\documentclass[twoside,letterpaper, onseide, final, conference, comsoc]{IEEEtran}
\usepackage[T1]{fontenc}
\usepackage[latin9]{inputenc}
\usepackage{geometry}
\usepackage{amsmath}
\usepackage{graphicx}

\makeatletter

\newcommand{\lyxdot}{.}

\IEEEoverridecommandlockouts \IEEEpubid{\makebox[\columnwidth]{ 978-1-5386-3531-5/17/\$31.00~\copyright~2017 IEEE \hfill} \hspace{\columnsep}\makebox[\columnwidth]{ }}
\setkeys{Gin}{width=\linewidth}
\usepackage{cite}

\long\def\@makecaption#1#2{\ifx\@captype\@IEEEtablestring%
\footnotesize\begin{center}{\normalfont\footnotesize #1}\\
{\normalfont\footnotesize\scshape #2}\end{center}%
\@IEEEtablecaptionsepspace
\else
\@IEEEfigurecaptionsepspace
\setbox\@tempboxa\hbox{\normalfont\footnotesize {#1.}~~ #2}%
\ifdim \wd\@tempboxa >\hsize%
\setbox\@tempboxa\hbox{\normalfont\footnotesize {#1.}~~ }%
\parbox[t]{\hsize}{\normalfont\footnotesize \noindent\unhbox\@tempboxa#2}%
\else
\hbox to\hsize{\normalfont\footnotesize\hfil\box\@tempboxa\hfil}\fi\fi}
\ifCLASSOPTIONcompsoc
  \usepackage[caption=false,font=normalsize,labelfont=sf,textfont=sf]{subfig}
\else
  \usepackage[caption=false,font=footnotesize]{subfig}
\fi
 
\usepackage{amsmath}
\DeclareMathOperator*{\argmax}{arg\,max}

\makeatother

\begin{document}

\title{The Impact of Adaptive Guards for~5G~and~Beyond}

\author{Ali Fatih Demir\IEEEauthorrefmark{1}, \IEEEmembership{Student Member, IEEE, }Hüseyin
Arslan\IEEEauthorrefmark{1}\IEEEauthorrefmark{2}\IEEEmembership{, Fellow, IEEE}\\\IEEEauthorblockA{\IEEEauthorrefmark{1}Department of Electrical Engineering, University
of South Florida, Tampa, FL, 33620}\IEEEauthorblockA{\IEEEauthorrefmark{2}School of Engineering and Natural Sciences,
Istanbul Medipol University, Istanbul, TURKEY, 34810}e-mail: afdemir@mail.usf.edu, arslan@usf.edu}
\maketitle
\begin{abstract}
The next generation communication systems are evolving towards an
increased flexibility in different aspects. Enhanced flexibility is
the key in order to address diverse requirements. This paper presents
the significance of adaptive guards considering a windowed-OFDM system
which supports a variety of services operating asynchronously under
the same network. The windowing approach requires a guard duration
to suppress the out-of-band emissions (OOBE), and the guard band is
required to handle the adjacent channel interference (ACI) along with
the windowing. The guards in both time and frequency domains are optimized
with respect to the use case and power offset between the users. To
fully exploit and further increase the potential of adaptive guards,
an interference-based scheduling algorithm is proposed as well. The
results show that the precise design that facilitates such flexibility
reduce the guards significantly and boost the spectral efficiency.
\end{abstract}



\section{Introduction}

The next generation communication systems including 5G are expected
to support high flexibility and a diverse range of services, unlike
the previous standards. The IMT-2020 vision defines the use cases
into three main categories as enhanced mobile broadband (eMBB), massive
machine type communications (mMTC), and ultra-reliable low-latency
communications (URLLC) featuring 20 Gb/s peak data rate, 10\textsuperscript{6}/km\textsuperscript{2}
device density, and less than 1 ms latency, respectively \cite{zhang2016}.
The applications which require larger bandwidth and spectral efficiency
fall into eMBB category, whereas the ones that have a tight requirement
for device battery life falls into mMTC. Usually, the industrial smart
sensors or medical implants \cite{demir2016a} need to operate several
years without maintenance, and hence low device complexity and high
energy efficiency are crucial for these mMTC services. Furthermore,
the mission-critical applications such as remote surgery \cite{demir2016b}
or self-driving vehicles are represented in URLLC. Therefore, a flexible
air interface is required to meet these different requirements. 

Orthogonal frequency-division multiplexing (OFDM) is the most popular
multi-carrier modulation scheme which is currently being deployed
in many standards such as 4G LTE and the IEEE 802.11 family \cite{hwang2009}.
A major disadvantage of OFDM systems is their high out-of-band emissions
(OOBE). The OFDM signal is well localized in the time domain with
a rectangular pulse shape, which corresponds to a sinc shape in the
frequency domain. The sidelobes of the sincs cause significant OOBE
and should be reduced to avoid adjacent channel interference (ACI).
Especially, the frequency localization is important to allow asynchronous
transmission across adjacent sub-bands and coexistence with other
waveforms/numerologies in the network \cite{demir2017a}. However,
a signal cannot be limited in both domains simultaneously due to the
Heisenberg\textquoteright s uncertainty principle \cite{benedicks1985}.
Hence, a better spectrum confinement is realized with the cost of
expansion in the time domain. Typically, OOBE is reduced by various
windowing/filtering approaches, and numerous waveforms are proposed
for the upcoming 5G standard to provide better time-frequency concentration
\cite{demir2017a,berardinelli2016,sahin2016,3gppQualCand,zhang2016,ankarali2017flexible}.
These filtering and windowing operations require additional period
which extends the guard duration between the consecutive OFDM symbols.
Also, extra guard bands are needed in between adjacent channels to
control the ACI along with the windowing/filtering that handles the
OOBE. The forthcoming generations must optimize the guards in both
time and frequency domains to boost the spectral efficiency. 

This paper presents the significance of adaptive guards considering
an OFDM-based system which supports a variety of services operating
asynchronously under the same network. The OOBE is reduced with a
transmitter windowing operation that smooths the inherent rectangular
pulse shape of OFDM. This technique retains the main design of the
OFDM receivers and provides backward compatibility for the existing
systems. The guard band and the window parameters that control the
guard duration are jointly optimized regarding the use case and the
power offset between the users. Although various windowing approaches
are proposed towards better spectral concentration \cite{weiss2004,bala2013,macaluso2014,guvenkaya2015},
this study also reduces the need for guards by grouping the users
with similar power levels and SIR requirements. Hence, the potential
of adaptive guards is further increased and exploited with an interference-based
scheduling algorithm.

The rest of the paper is organized as follows. Section \ref{sec:II}
describes the system model and explains the methodology in detail.
Section \ref{sec:III} presents optimization of guards regarding the
user requirements. Section \ref{sec:IV} proposes the interference-based
scheduling strategy along with the utilization of adaptive guards.
Finally, Section \ref{sec:V} summarizes the contributions and concludes
the paper. 

\section{System Model\label{sec:II}}

\begin{figure}[b]
\centering\includegraphics[width=1\columnwidth]{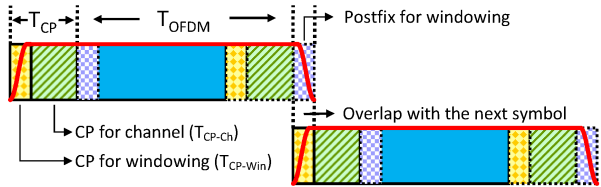}
\centering{}\caption{Transmitter windowing operation and the guard durations.\label{fig:W-OFDM}}
\end{figure}

Consider a multiuser OFDM-based system where asynchronous numerologies
operate in the network. The users which have different use cases (i.e.,
requirements) and power levels perform transmitter windowing to control
their OOBE levels and reduce interference to the users serving in
adjacent bands. The guard duration that is designated for the multi-path
channel is fixed and sufficient to handle the inter-symbol interference
(ISI). An additional guard duration is required to perform windowing.
Several windowing functions have been evaluated in detail \cite{farhang2011}
with different tradeoffs between the width of the main lobe and suppression
of the side lobes. The optimal windowing function is beyond the scope
of this study, and the raised-cosine (RC) window is adopted due to
its computational simplicity and common use in the literature \cite{weiss2004,bala2013,macaluso2014}.
The RC window function \cite{weiss2004} is expressed as follows:

\begin{equation}
\hspace{-8.5pt}g[n]=\begin{cases}
\frac{1}{2}+\frac{1}{2}\cos\left(\pi+\frac{\pi n}{\alpha N_{\textrm{T}}}\right) & 0\leq n\leq\alpha N_{\textrm{T}}\\
1 & \alpha N_{\textrm{T}}\leq n\leq N_{\textrm{T}}\\
\frac{1}{2}+\frac{1}{2}\cos\left(\pi-\frac{\pi n}{\alpha N_{\textrm{T}}}\right) & N_{\textrm{T}}\leq n\leq\left(\alpha+1\right)N_{\textrm{T}}
\end{cases}\label{eq:Eq1-WindowRC}
\end{equation}
where $\alpha$ is the roll-off factor ($0\leq\alpha\leq1$) and $N_{\textrm{T}}$
is the symbol length of the RC function. The roll-off factor ($\alpha$)
controls the taper duration of the window. As $\alpha$ increases,
the OOBE decreases at the price of increased guard duration to perform
windowing. The transmitter windowing operation is illustrated in Fig.
\ref{fig:W-OFDM}. First, the cyclic prefix (CP) that is allocated
to deal with the multi-path channel is further extended on both edges,
and then the extended part from the beginning of the OFDM symbol is
appended to the end. The transitions parts (i.e., ramp-ups and ramp-downs)
of adjacent symbols are overlapped to decrease the additional time-domain
overhead resulting from the windowing operation.

\begin{figure}[b]
\centering\includegraphics[width=0.69\columnwidth]{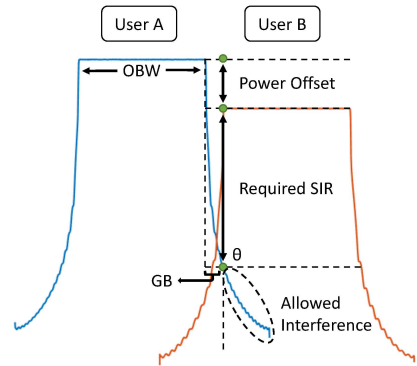}
\centering{}\caption{Guard band allocation considering the interference threshold ($\theta$)
in the adjacent band. (OBW stands for occupied bandwidth and $\theta$
is represented in dB w.r.t. the power of ``user A'' throughout the
paper)\label{fig:IntThresh}}
\end{figure}

The windowing operation is not sufficient to handle the OOBE, and
non-negligible guard bands are still needed. However, the amount of
guard band (GB) or the length of guard duration (GD) to perform windowing
depends on the power offset and the required signal to interference
ratio (SIR) level of the users in adjacent bands. As an example, the
leaked energy from the near user can be more powerful than the in-band
energy of the far user in its adjacent band (i.e., the well-known
near-far problem). The power control mechanism is a solution for this
power offset problem. Nonetheless, it prevents near users to deploy
higher order modulation schemes. Thus, the power control needs to
be relaxed with an adaptive design to improve the spectral efficiency. 

The required guards in both time and frequency domains are tightly
related to the use case as well. For example, the guard units and
other extra overheads decrease the spectral efficiency, which is especially
critical for the eMBB type of communications. Hence, the guards are
reduced with an expense of interference on the adjacent bands. On
the other hand, the reliability and latency are extremely important
for mission critical communications where errors and retransmissions
are less tolerable. Thus, a strict OOBE suppression is more feasible
for URLLC applications. In addition, mMTC operates at the low power
level to preserve energy and might suffer from the ACI seriously in
an asynchronous heterogeneous network.

As a result of these discussions, the threshold for allowed interference
level ($\theta$) on adjacent bands should be adaptive considering
the power offset (PO) and the use case. Fig. \ref{fig:IntThresh}
presents how GB is inserted regarding $\theta$ to achieve the desired
SIR level when there is a PO between the users in adjacent bands.
Throughout the numerical evaluations in the paper, $T_{CP-Win}$ (i.e.,
GD) and GB are adaptive, and these guards are optimized in Section
\ref{sec:III}. The rest of the parameters belong to the windowed-OFDM
(W-OFDM) system are fixed and summarized in Table \ref{tab:Sim-Par}.

\begin{table}[!t]
\caption{Simulation Parameters \label{tab:Sim-Par}}

\centering\includegraphics[width=0.55\columnwidth]{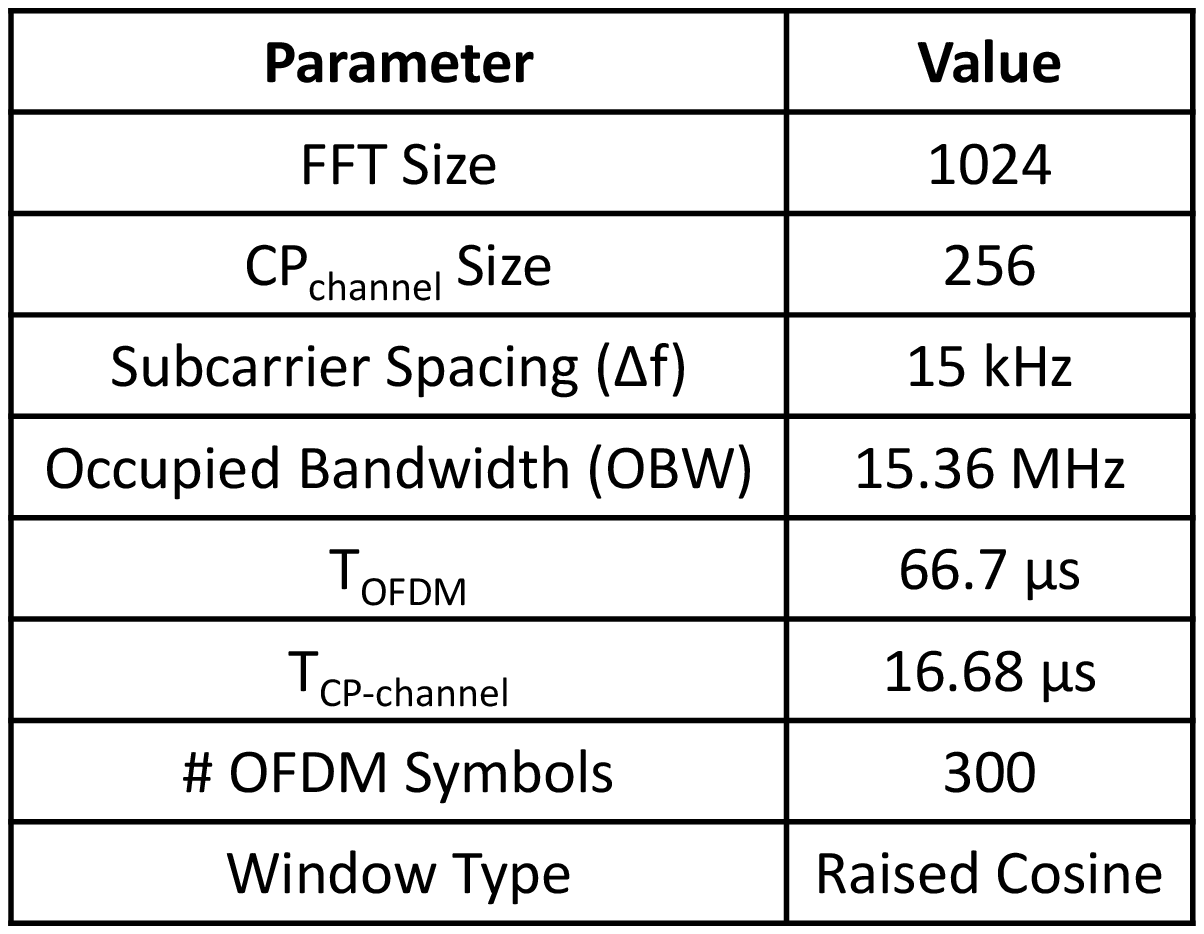}
\end{table}

\section{Optimization of The Adaptive Guards\label{sec:III}}

\begin{figure}[b]
\centering\includegraphics[width=1\columnwidth]{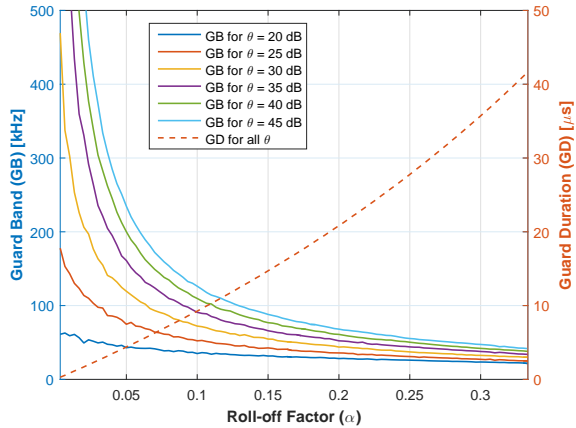}
\centering{}\caption{Required GB and GD to achieve selected $\theta$ levels.\label{fig:GBvsGD}}
\end{figure}

The ACI is handled by windowing and allocating guard band between
adjacent users as described in Section \ref{sec:II}. Since windowing
operation suppresses the OOBE with a penalty of additional guard duration,
the procedure converges to the utilization of guard duration (GD)
and guard band (GB) to achieve desired interference threshold ($\theta$).
Fig.\ref{fig:GBvsGD} presents the required GB and GD for selected
$\theta$. Each $\alpha$ value in the figure corresponds to a GD
to perform windowing, and a GB to deal with the remaining interference
power for a given $\theta$. 

\begin{figure}[t]
\centering\includegraphics[width=1\columnwidth]{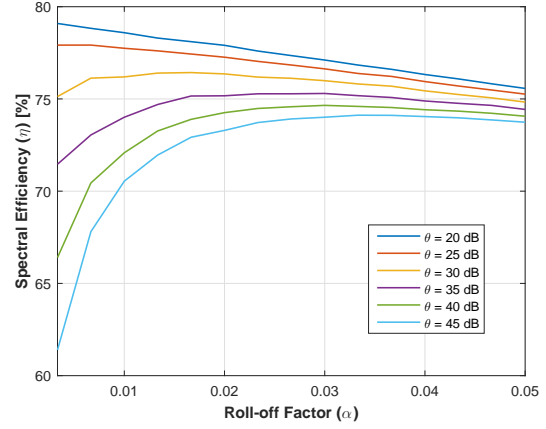}
\centering{}\caption{Spectral efficiency ($\eta$) of the GB and GD pairs that achieves
selected $\theta$. (Please note that each $\alpha$ corresponds to
a GB-GD pair as shown in Fig. \ref{fig:GBvsGD})\label{fig:SpecEff}}
\end{figure}

\begin{table}[b]
\caption{The optimal guards for selected $\theta$ \label{tab:OOBEthresh}}

\centering\includegraphics[width=1\columnwidth]{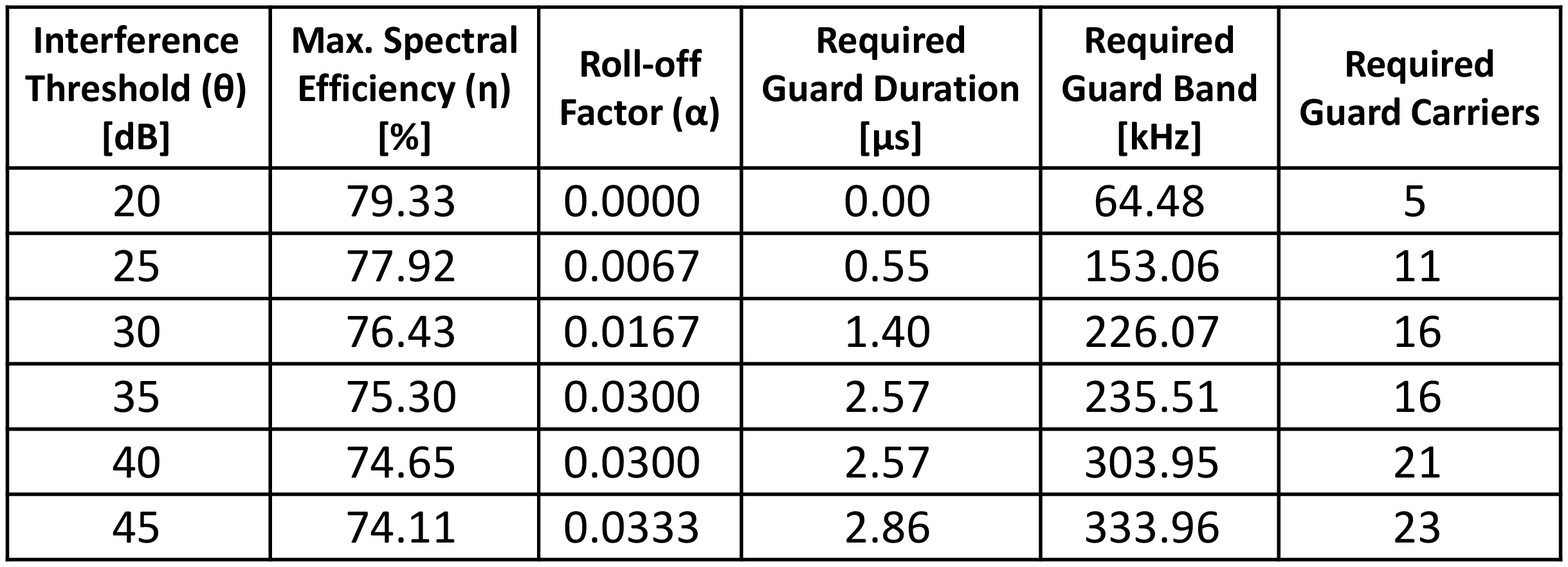}
\end{table}

An excessive amount of resources is needed to solve the problem only
with GB or GD. As a result, the spectral efficiency, which is defined
as the information rate that can be transmitted over a given bandwidth,
decreases significantly. Therefore, GB and GD must be jointly optimized
to boost the efficiency of the communication system. This hyper-parameter
optimization has been performed by a grid search algorithm through
a manually specified subset of the hyper-parameter space \cite{bergstra2011}.
The spectral efficiency ($\eta$) is proportional to the multiplication
of efficiencies in the time and frequency domains which are expressed
as follows:

\begin{equation}
\eta_{time}=\frac{T_{OFDM}}{T_{OFDM}+T_{CP-Ch}+T_{CP-Win}}\label{eq:Eq2-nTime}
\end{equation}

\begin{equation}
\eta_{freq}=\frac{OBW}{OBW+(GB\times2)}\label{eq:Eq3-nFreq}
\end{equation}
Since $T_{OFDM}$, $T_{CP-Ch}$, and $OBW$ are fixed parameters,
the degrees of freedom that can be selected independently becomes
only $T_{CP-Win}$ (i.e., GB) and GD. The problem that seeks for the
optimal GB and GD pair can be formulated as follows: 

\begin{equation}
(GB,\:GD)\:=\argmax_{GB,\:GD}(\eta_{time}\times\eta_{freq})\:,\label{eq:Eq4-Objective}
\end{equation}

\vspace{-5pt}

\begin{equation}
\begin{array}{cc}
\textrm{subject\:to:} & PO+SIR\leq\theta\:.\end{array}\label{eq:Eq5-Constraint}
\end{equation}

The spectral efficiency of the W-OFDM system for selected $\theta$
values is presented in Fig. \ref{fig:SpecEff}. Each $\alpha$ value
in the graph corresponds to a GB-GD pair for a given $\theta$ and
the peak value of each curve provide the optimal pair. These optimal
pairs are listed in Table \ref{tab:OOBEthresh} along with the associated
parameters. The results show that the need for windowing decreases
as $\theta$ decreases, and hence the desired ACI level can be achieved
only with a few guard carriers. In addition, the spectral efficiency
increases with the decrease in $\theta$. The variation in required
guards clearly affirms that the adaptive design improves the spectral
efficiency significantly instead of designing the system considering
the worst case (e.g., $\eta_{\theta=45\:dB}$ = 74.11 \% whereas $\eta_{\theta=20\:dB}$
= 79.33 \%). 

\section{Interference-based Scheduling\label{sec:IV}}

The optimization results in Section \ref{sec:III} reveal that the
spectral efficiency ($\eta$) decreases as the interference threshold
($\theta$) increases. Since $\theta$ depends on the users operating
in the adjacent bands, the potential of adaptive guards can be increased
further along with the utilization of an interference-based scheduling
algorithm. Assuming that the base station or the user equipment has
all necessary information, $\theta$ is determined as follows: 

\begin{equation}
\theta_{i}=max(SIR_{i-1}+PO_{i-1},SIR_{i+1}+PO_{i+1}),\:\:\:\:\:\:\forall i\label{eq:Eq4-ThetaSchedule}
\end{equation}
where $i$ is the indicator of the available consecutive bands. If
the users with similar power levels and SIR requirements are grouped
together, the average $\theta$ in the network decreases. As a result,
the need for guards is reduced, and the spectral efficiency increases. 

\begin{figure}[b]
\includegraphics[width=1\columnwidth]{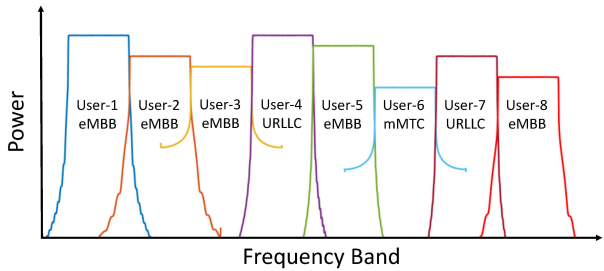}

\caption{Random scheduling of eight users which have different requirements.\label{fig:RandomBands}}
\end{figure}

\begin{figure}[b]
\includegraphics[width=1\columnwidth]{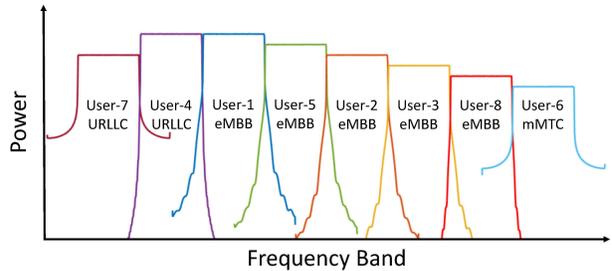}

\caption{Interference-based scheduling of eight users which have different
requirements.\label{fig:ScheduledBands}}
\end{figure}

Consider an exemplary scenario with eight users, where the users have
different power levels and SIR requirements as shown in Tables \ref{tab:RandomBands}
and \ref{tab:ScheduledBands}. The power offset (PO) pairs in the
tables are provided regarding the users in adjacent bands. The users
are assigned to the bands in two different ways. In the first scenario,
a random scheduling has been realized (Fig. \ref{fig:RandomBands}),
whereas the ACI based scheduling strategy is utilized in the second
scenario (Fig. \ref{fig:ScheduledBands}). To compare and present
the impact of the adaptive guards, a fixed guard assignment strategy
is implemented in the random scheduling scenario as well. The guards
are selected regarding the worst case scenario (i.e., $\theta$ =
45 dB) in the fixed assignment scenario. 

The comparison of required guards for the fixed guard assignment with
random scheduling, the adaptive guard assignment with random scheduling,
and the adaptive guard assignment with interference-based scheduling
scenarios is summarized in Table \ref{tab:Comp}. The results show
that the amount of guard duration (GD) and guard band (GB) decreased
by 57\% and 19\%, respectively when the fixed guards are replaced
with the adaptive guards. In addition, the amount of GD and GB decreased
further by 35\% and 16\%, respectively when the random scheduling
is replaced with the interference-based scheduling strategy. 

\section{Conclusions\label{sec:V}}

This paper presented the importance of adaptive guards considering
a windowed-OFDM system which supports a variety of services operating
asynchronously under the same network. The guards in both time and
frequency domains are optimized taking the use case and power offset
into account. Furthermore, the need for guards is reduced with an
interference-based scheduling algorithm. Such a scheduling strategy
is especially critical when the power offset between the users operating
in adjacent bands is high. The results show that the precise design
that facilitates such flexibility considering the user requirements
improve the spectral efficiency significantly. 
\begin{table}[!t]
\caption{The requirements of randomly scheduled users\label{tab:RandomBands}}

\centering\includegraphics[width=1\columnwidth]{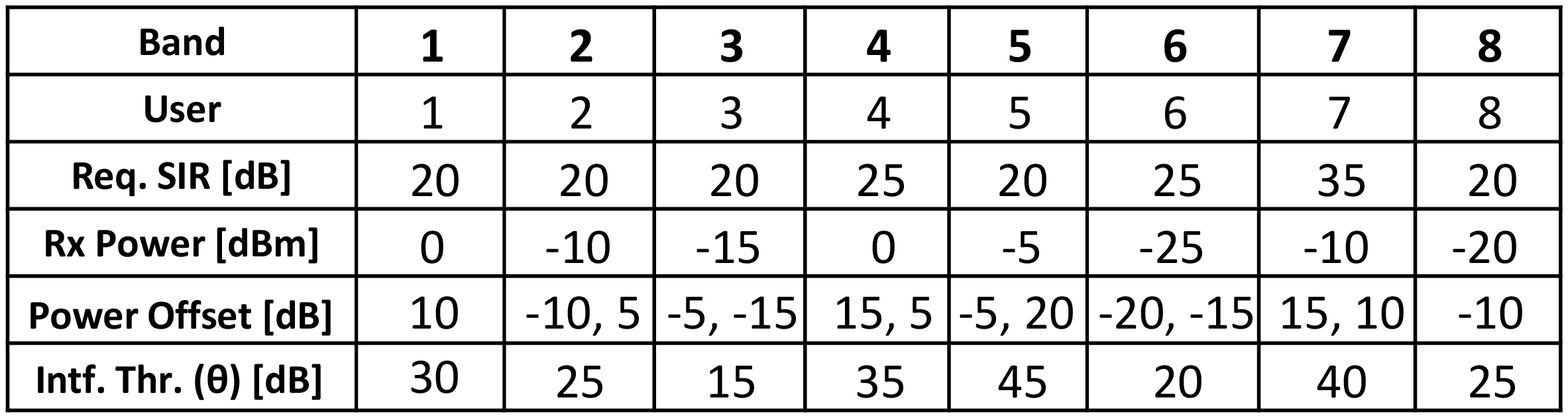}
\end{table}
\begin{table}[!t]
\caption{The requirements of interferece-based scheduled users\label{tab:ScheduledBands}}

\centering\includegraphics[width=1\columnwidth]{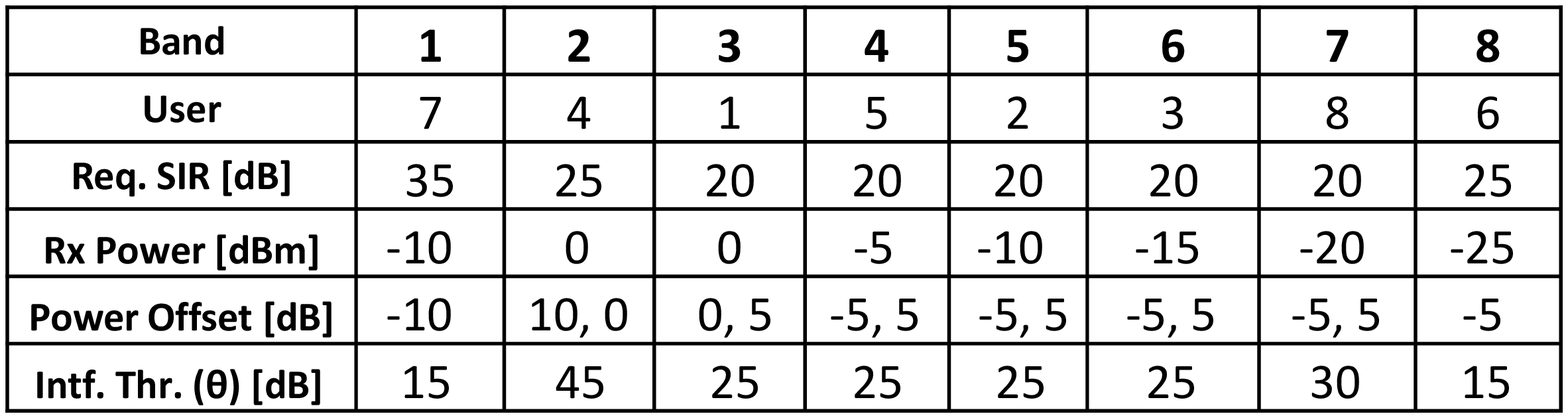}
\end{table}
\begin{table}[!t]
\caption{The comparison of required guards for different scenarios\label{tab:Comp}}

\centering\includegraphics[width=0.75\columnwidth]{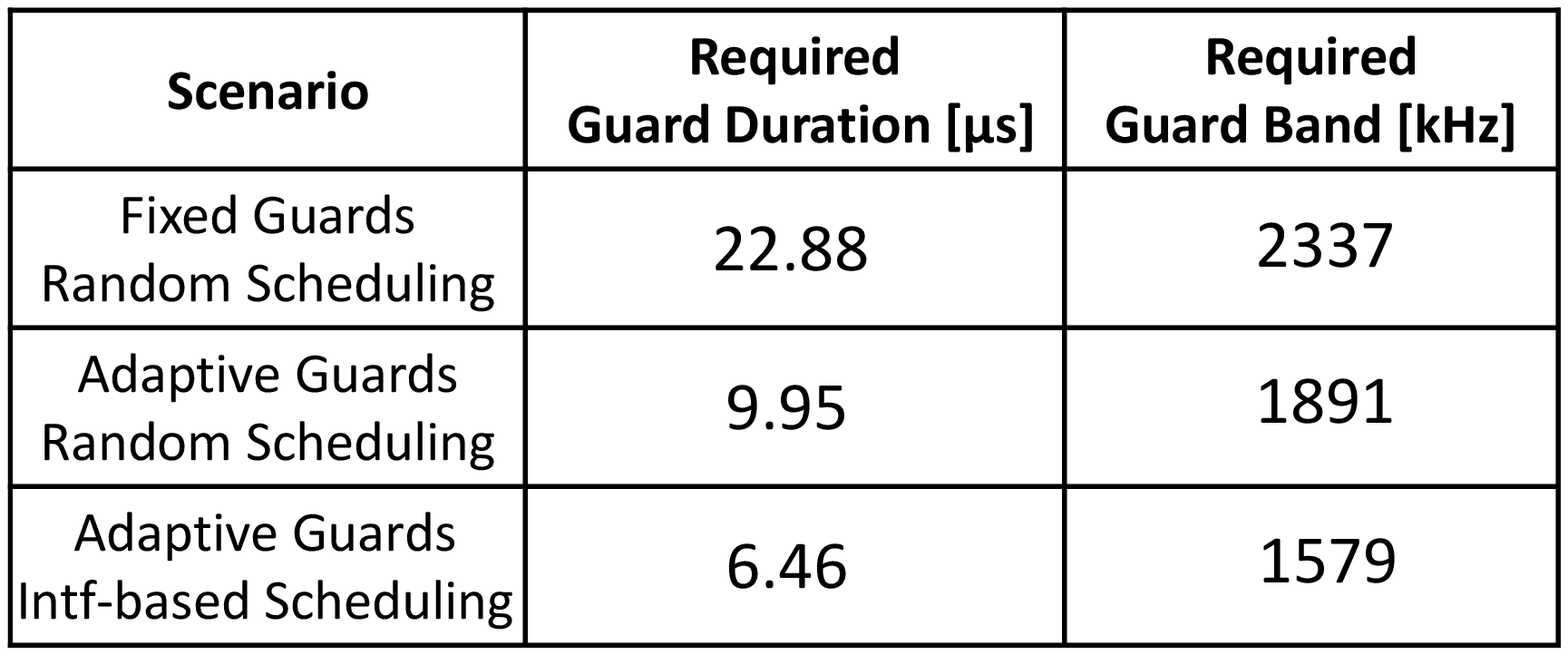}
\end{table}
Although the computational complexity increases compared to conventional
OFDM-based systems, the computation of the optimal GB and GD is an
offline process that requires a one-time solution. Hence, a lookup
table method can be adopted to reduce the complexity. This study will
be extended by optimizing the guards and scheduling the users under
various channel conditions and impairments. Also, the proposed methodology
is applicable to the filtered-OFDM systems. The next generation communications
systems are evolving towards an increased flexibility in different
aspects. Enhanced flexibility is the key especially to address diverse
requirements, and definitely, the guards should be a part of the flexibility
consideration as well.

\bibliographystyle{IEEEtran}
\bibliography{Waveform5GRef}

\begin{thebibliography}{10}
\providecommand{\url}[1]{#1}
\csname url@samestyle\endcsname
\providecommand{\newblock}{\relax}
\providecommand{\bibinfo}[2]{#2}
\providecommand{\BIBentrySTDinterwordspacing}{\spaceskip=0pt\relax}
\providecommand{\BIBentryALTinterwordstretchfactor}{4}
\providecommand{\BIBentryALTinterwordspacing}{\spaceskip=\fontdimen2\font plus
\BIBentryALTinterwordstretchfactor\fontdimen3\font minus
  \fontdimen4\font\relax}
\providecommand{\BIBforeignlanguage}[2]{{%
\expandafter\ifx\csname l@#1\endcsname\relax
\typeout{** WARNING: IEEEtran.bst: No hyphenation pattern has been}%
\typeout{** loaded for the language `#1'. Using the pattern for}%
\typeout{** the default language instead.}%
\else
\language=\csname l@#1\endcsname
\fi
#2}}
\providecommand{\BIBdecl}{\relax}
\BIBdecl

\bibitem{zhang2016}
X.~Zhang, L.~Chen, J.~Qiu, and J.~Abdoli, ``On the {Waveform} for 5{G},''
  \emph{IEEE Communications Magazine}, vol.~54, no.~11, pp. 74--80, Nov. 2016.

\bibitem{demir2016a}
A.~F. Demir, Z.~E. Ankarali, Q.~H. Abbasi, Y.~Liu, K.~Qaraqe, E.~Serpedin,
  H.~Arslan, and R.~D. Gitlin, ``In {Vivo} {Communications}: {Steps} {Toward}
  the {Next} {Generation} of {Implantable} {Devices},'' \emph{IEEE Vehicular
  Technology Magazine}, vol.~11, no.~2, pp. 32--42, Jun. 2016.

\bibitem{demir2016b}
A.~F. Demir, Q.~Abbasi, Z.~E. Ankarali, A.~Alomainy, K.~Qaraqe, E.~Serpedin,
  and H.~Arslan, ``Anatomical {Region}-{Specific} {In} {Vivo} {Wireless}
  {Communication} {Channel} {Characterization},'' \emph{IEEE Journal of
  Biomedical and Health Informatics}, vol.~PP, no.~99, pp. 1--1, 2016.

\bibitem{hwang2009}
T.~Hwang, C.~Yang, G.~Wu, S.~Li, and G.~Y. Li, ``{OFDM} and {Its} {Wireless}
  {Applications}: {A} {Survey},'' \emph{IEEE Transactions on Vehicular
  Technology}, vol.~58, no.~4, pp. 1673--1694, May 2009.

\bibitem{demir2017a}
A.~F. Demir, M.~Elkourdi, M.~Ibrahim, and H.~Arslan, ``{W}aveform {D}esign for
  5{G} and {B}eyond,'' accepted for publication in "5G Networks: Fundamental
  Requirements, Enabling Technologies, and Operations Management". {John Wiley
  \& Sons, Ltd}, 2017.

\bibitem{benedicks1985}
M.~Benedicks, ``On {Fourier} transforms of functions supported on sets of
  finite {Lebesgue} measure,'' \emph{Journal of Mathematical Analysis and
  Applications}, vol. 106, no.~1, pp. 180--183, Feb. 1985.

\bibitem{berardinelli2016}
G.~Berardinelli, K.~I. Pedersen, T.~B. Sorensen, and P.~Mogensen, ``Generalized
  {DFT}-{Spread}-{OFDM} as 5{G} {Waveform},'' \emph{IEEE Communications
  Magazine}, vol.~54, no.~11, pp. 99--105, Nov. 2016.

\bibitem{sahin2016}
A.~Sahin, R.~Yang, E.~Bala, M.~C. Beluri, and R.~L. Olesen, ``Flexible
  {DFT}-{S}-{OFDM}: {Solutions} and {Challenges},'' \emph{IEEE Communications
  Magazine}, vol.~54, no.~11, pp. 106--112, Nov. 2016.

\bibitem{3gppQualCand}
{Q}ualcomm {I}nc., ``{W}aveform {C}andidates,'' 3{GPP} {S}tandard
  {C}ontribution ({R}1-162199), {B}usan, {K}orea, {A}pr. 11-15 2016.

\bibitem{ankarali2017flexible}
Z.~E. {Ankarali}, B.~{Peköz}, and H.~{Arslan}, ``Flexible {Radio} {Access}
  {Beyond} {5G}: {A} {F}uture {P}rojection on {W}aveform, {N}umerology, and
  {F}rame {D}esign {P}rinciples,'' \emph{IEEE Access}, vol.~5, pp.
  18\,295--18\,309, 2017.

\bibitem{weiss2004}
T.~Weiss, J.~Hillenbrand, A.~Krohn, and F.~K. Jondral, ``Mutual interference in
  {OFDM}-based spectrum pooling systems,'' in \emph{2004 {IEEE} 59th
  {Vehicular} {Technology} {Conference}. {VTC} 2004-{Spring} ({IEEE}}, vol.~4,
  May 2004, pp. 1873--1877 Vol.4.

\bibitem{bala2013}
E.~Bala, J.~Li, and R.~Yang, ``Shaping {Spectral} {Leakage}: {A} {Novel}
  {Low}-{Complexity} {Transceiver} {Architecture} for {Cognitive} {Radio},''
  \emph{IEEE Vehicular Tech. Mag.}, vol.~8, no.~3, pp. 38--46, Sep. 2013.

\bibitem{macaluso2014}
I.~Macaluso, B.~Ozgul, T.~K. Forde, P.~Sutton, and L.~Doyle, ``Spectrum and
  {Energy} {Efficient} {Block} {Edge} {Mask}-{Compliant} {Waveforms} for
  {Dynamic} {Environments},'' \emph{IEEE Journal on Selected Areas in
  Communications}, vol.~32, no.~2, pp. 307--321, Feb. 2014.

\bibitem{guvenkaya2015}
E.~G{\"u}venkaya, A.~{\c S}ahin, E.~Bala, R.~Yang, and H.~Arslan, ``A
  {Windowing} {Technique} for {Optimal} {Time}-{Frequency} {Concentration} and
  {ACI} {Rejection} in {OFDM}-{Based} {Systems},'' \emph{IEEE Transactions on
  Communications}, vol.~63, no.~12, pp. 4977--4989, Dec. 2015.

\bibitem{farhang2011}
B.~Farhang-Boroujeny, ``{OFDM} {Versus} {Filter} {Bank} {Multicarrier},''
  \emph{IEEE Signal Processing Mag.}, vol.~28, no.~3, pp. 92--112, May 2011.

\bibitem{bergstra2011}
J.~S. Bergstra, R.~Bardenet, Y.~Bengio, and B.~K{\'e}gl, ``Algorithms for
  {Hyper}-{Parameter} {Optimization},'' in \emph{Advances in {Neural}
  {Information} {Processing} {Systems} 24}, J.~Shawe-Taylor, R.~S. Zemel, P.~L.
  Bartlett, F.~Pereira, and K.~Q. Weinberger, Eds.\hskip 1em plus 0.5em minus
  0.4em\relax Curran Associates, Inc., 2011, pp. 2546--2554.

\end{thebibliography}

\end{document}